\documentclass[preprint]{aastex}

\usepackage{epsfig}
\usepackage{amsmath}
\usepackage{amssymb}
\usepackage{appendix}

\def\be{\begin{equation}}
\def\ee{\end{equation}}
\def\ba{\begin{eqnarray}}
\def\ea{\end{eqnarray}}
\def\go{\mathrel{\raise.3ex\hbox{$>$}\mkern-14mu
             \lower0.6ex\hbox{$\sim$}}}
\def\lo{\mathrel{\raise.3ex\hbox{$<$}\mkern-14mu
             \lower0.6ex\hbox{$\sim$}}}

\begin{document}

\title{Tidal Novae in Compact Binary White Dwarfs}

\author{Jim Fuller and Dong Lai}
\affil{Department of Astronomy, Cornell University,
    Ithaca, NY 14850}

\begin{abstract}

Compact binary white dwarfs (WDs) undergoing orbital decay due to
gravitational radiation can experience significant tidal heating prior
to merger. In these WDs, the dominant tidal effect involves the
excitation of outgoing gravity waves in the inner stellar envelope and
the dissipation of these waves in the outer envelope. As
the binary orbit decays, the WDs are synchronized from outside in
(with the envelope synchronized first, followed by the core).  We
examine the deposition of tidal heat in the envelope of a Carbon-Oxygen
WD and study how such tidal heating affects the structure and evolution of the WD.
We show that significant tidal heating can occur in the
star's degenerate hydrogen layer. This layer heats up faster than it cools, triggering runaway nuclear fusion. Such ``tidal novae'' may occur in all WD binaries containing a CO WD, at orbital periods
between 5~min and 20~min, and precede the final merger by $10^5$-$10^6$~years. 

\end{abstract}

\section{Introduction}

Compact white dwarf (WD) binaries (with orbital periods in the range
from minutes to hours) are important for several areas of
astrophysics. The orbits of these systems decay via the emission of
gravitational waves, constituting the largest signals for the next
generation space-based gravitational wave interferometers. Systems of
sufficiently short orbital period will merge within a Hubble time, the
result of which may produce a variety of exotic objects, such as
helium-rich sd0 stars, R CrB stars and AM CVn binaries.  Most
importantly, when the total binary mass is near the Chandrasekhar
limit, the merged WDs may collapse into a neutron star or explode as a Type Ia supernova
(e.g., Webbink 1984; Iben \& Tutukov 1984).
Recent studies have provided support for such ``double degenerate''
progenitors of SNe Ia.  (e.g., Gilfanov \& Bogdan 2010; Di Stefano
2010; Maoz et al.~2010; Li et al.~2011; Bloom et al.~2012;
Schaefer \& Pagnotta 2012).

The outcome of a WD binary merger depends on the masses of the WDs and
their pre-merger conditions (e.g., Segretain et al.~1997; Yoon et al.~2007;
Loren-Aguilar et al.~2009; van Kerkwijk et al.~2010; Dan et al.~2012;
Raskin et al.~2012). Most previous studies of pre-merger binary WDs have 
focused equilibrium tides and considered tidal dissipation in a 
perameterized way (e.g., Mochkovitch \& Livio 1989; Iben et al. 1998; 
Willems et al.~2010; Piro 2011). None of these studies have sought to 
predict the magnitude and location of tidal heating due to dynamical tides,
which dominate the tidal responses of the binary WDs.

In two recent papers (Fuller \& Lai 2011, 2012, hereafter paper I and paper
II, respectively), we presented the first {\it ab initio}
calculations of dynamical tides in realistic WD models. In paper I, we
considered resonant excitations of WD g-modes during binary decay and
showed that the modes reach non-linear amplitudes near
the surface of the star. 
This implies that, rather than exciting discrete g-modes, the binary
companion will excite a continuous train of gravity waves, which
propagate outward and dissipate in the outer envelope of the WD.  We
studied such continuous tidally excited waves in paper II.  For a
canonical Carbon/Oxygen WD (consisting of CO core with a He-H
envelope), we showed that the outgoing waves are primarily launched
at the CO/He transition region, and propagate toward the WD surface,
where they are likely dissipated through a combination of non-linear
processes and radiative damping.
We computed the energy and angular momentum
flux carried by the waves in order to predict the orbital and spin
evolution of WDs in compact binaries. We found that such dynamical tides
cause the binary WDs to be nearly synchronized prior to merger. Furthermore, the
tidal heating rate can be quite large at short orbital periods (exceeding
tens of solar luminosities just before merger, depending on the system
parameters), potentially leading to significant observable signatures.

In this {\it Letter}, we show that tidal heating may 
trigger a thermonuclear runaway hydrogen fusion event in a CO WD. 
The observational consequence of such an event would likely be an outburst
that resembles a classical nova. We call this new phenomenon a ``Tidal Nova'' (TN).
Unlike all other types of novae or supernovae, a TN does not rely on
mass accretion or collapse.
We present a simple two-zone model for the angular
momentum evolution of a differentially rotating WD, which we use to calculate 
the radial tidal heating profile within the WD. 
We then evolve the WD model including tidal heating to 
calculate changes in its temperature, luminosity, and
internal structure. For a wide range of physically plausible
parameters, we demonstrate that tidal heating induces a
thermonuclear runaway event. Finally, we discuss the observational
signatures of such an event, and compare our predictions to
observations of short-period WD binaries.

\section{Energy and Angular Momentum of Tidally Excited Gravity Waves}

Using the method described in Paper II, we calculate the amplitude of
tidally excited gravity waves inside a WD.
We consider a circular orbit with angular frequency $\Omega$.
The WD spins at an angular frequency $\Omega_s$, and the spin is 
aligned with the orbit. In the corotating frame, the frequency of the dominant $l=m=2$ tidal potential
is $\omega=2(\Omega-\Omega_s)$.
For a WD of mass $M$ and radius $R$ (and given internal structure)
with a companion of mass $M'$, the energy and angular momentum 
fluxes carried by the gravity waves can be written as
\ba
&&\dot{J}_z(\Omega,\omega) = T_0(\Omega) F(\omega),\label{Jdot}\\
&& \dot{E}(\Omega,\omega) = \Omega T_0 F(\omega),\label{Edot}
\ea
where
\be
\label{T0}
T_0(\Omega) = \frac{G M'^2}{a} \bigg(\frac{R}{a}\bigg)^5,
\ee
with $\Omega=\sqrt{GM_t/a^3}$ the orbital angular
frequency ($M_t=M+M'$ is the total mass and $a$ is the orbital semi-major axis). 

The dimensionless function $F(\omega)$ (similar to the tidal lag angle
in the language of equilibrium tide theory) determines the magnitude of wave
excitation, and is strongly dependent on the internal
structure of the WD and the tidal frequency $\omega$. In Paper II we
have calculated $F(\omega)$ for $0.6 M_\odot$ CO WD models 
of various surface temperatures and slow rotation.
We found that $F(\omega)$ is an erratic function of $\omega$
because of the ``quasi-resonance cavity'' formed by the CO core inside 
the He/H shell. However, because of the strong dependence of $F(\omega)$
on $\omega$ [the envelope of $F(\omega)$ approximately scales as $\omega^5$], 
at sufficiently short orbital periods, tidal spin-up
combined with orbital decay via gravitational radiation ensure that
$\omega \simeq {\rm const}$. The orbital period at which this
transition occurs is $P_c \simeq  40$ minutes, depending on the WD
masses and temperatures [see Eq.~(79) of Paper II]. 
At periods $P \lesssim P_c$, the tidal energy transfer rate is
\be
\label{Edot2}
\dot{E} \simeq \frac{3I\Omega^2}{2t_{\rm GW}},
\ee
where $I$ is the moment of inertia of the WD, 
and $t_{\rm GW}=|a/\dot a|$ is the binary inspiral time due to gravitational radiation, 
\be
t_{\rm GW}= 4.2\times 10^{5}\,{\rm yr} \bigg(\frac{M_{\odot}^2}{MM'}\bigg)\bigg(\frac{M_t}
{2M_{\odot}}\bigg)^{\!\!1/3}\!\! \bigg(\!\frac{P}{10\,{\rm min}}
\bigg)^{\!\! 8/3}.
\label{tgw}\ee

When the outgoing gravity waves damp in the WD envelope and locally
deposit their angular momentum, some of the wave energy is
converted into rotational kinetic energy, while the rest is converted
to heat. The heating rate is 
\be
\label{eheat}
\dot{E}_{\rm heat} = \dot E \bigg(1 - \frac{\Omega_{s}}{\Omega}\bigg).
\ee
If the WD maintains some differential rotation, $\Omega_s$ in the
above equation should be the rotation rate of the layer in which the
waves damp, and heat will also be generated through viscous angular
momentum transport.

\section{Two Zone Model for Tidal Heat Depostion}
\label{critical}

Our calculations indicate that the gravity waves reach non-linear
amplitudes and break in the outer layers of the WD. The location of
wave breaking depends on various parameters (e.g., orbital and tidal
frequencies), but is always at $r \gtrsim 0.92R$ and the exterior mass
$\Delta M \lesssim 10^{-4} M$ (Paper II).
Since a small fraction of the stellar mass absorbs the entire angular 
momentum flux, the outer layer spins up rapidly. 
If it spins up faster than angular momentum can be transported to the core, 
the outer layer will rotate synchronously with the
orbit. Outgoing waves approaching the synchronized envelope will be
absorbed near corotation and deposit their angular momentum,
causing the synchronized envelope to move to larger depths
(see Goldreich \& Nicholson 1989).

We consider a simple two-zone model for the spin evolution of the WD.
In this model, the envelope of the star rotates synchronously
with the orbit ($\Omega_{\rm env} = \Omega$), while the core 
rotates sub-synchronously ($\Omega_{\rm core} < \Omega$). The
envelope and core are coupled, with angular momentum being transferred
to the core according to a parameterized coupling time, 
$t_{\rm coup}$. The angular momentum of the core-envelope 
system evolves according to
\ba
&&\label{Jedot}
\frac{d}{dt}\left(I_{\rm env}\Omega_{\rm env}\right)
= \dot{J}_z(\Omega,\omega_{\rm core}) - \frac{I_{\rm env}}{t_{\rm coup}} (\Omega_{\rm env} - \Omega_{\rm core}),\\
&&\label{Jcdot} 
\frac{d}{dt}\left(I_{\rm core}\Omega_{\rm core}\right)
= \frac{I_{\rm env}}{t_{\rm coup}} (\Omega_{\rm env} - \Omega_{\rm core}),
\ea
where $I_{\rm env}=I-I_{\rm core}$ is the moment of inertia of the
envelope.  Here, $\dot{J}_z$ is the angular momentum flux which can be
calculated from equation (\ref{Jdot}). We have assumed that the
gravity waves are excited in the core and absorbed in the
envelope\footnote{This assumption is valid as long as long as the
core-envelope boundary is above the C/He transition layer (with an
exterior mass $\Delta M \approx 10^{-2} M_\odot$), which is the
region where the outgoing gravity waves are excited.}. 
Consequently, the angular momentum source term
$\dot{J}_z$ is only present in the envelope evolution equation,
although it is dependent on the tidal frequency in the core,
$\omega_{\rm core} = 2(\Omega - \Omega_{\rm core})$. 
Using $\Omega_{\rm env} =\Omega$, equations (\ref{Jedot}) and (\ref{Jcdot}) can be integrated to find
$I_{\rm env}$ and $\Omega_{\rm core}$ as a function of time or orbital period. The mass $\Delta 
M_{\rm env}$ of the envelope corresponds to $I_{\rm env} \simeq (2/3)\Delta M_{\rm env}R^2$. 

The thickness (or $\Delta M_{\rm env}$) of the envelope is dependent
on the parameter $t_{\rm coup}$. In stably stratified stars like WDs,
angular momentum can be transported by magnetic fields. In the presence
of a poloidal field $B$ connecting the core and envelope, $t_{\rm coup}$ 
can be estimated from the Alfven wave crossing time,
\be
\label{ta}
t_A = \frac{R\sqrt{4\pi\rho}}{B} \simeq 10^2 \ {\rm yr} \bigg(\frac{10^3 {\rm G}}{B}\bigg)
\ee
for our CO WD model. For WDs without an intrinsic magnetic field,
angular momentum may be transported via the Tayler-Spruit dynamo
(Spruit 2002). To estimate $t_{\rm coup}$, we calculate
the effective viscosity for angular momentum transport via the
Tayler-Spruit dynamo, $\nu_{TS}$, as outlined in Spruit 2002.\footnote{For simplicity, we have calculated the viscosity $\nu_{TS}$ without including the effects of composition gradients in the WD [see equation (32) in Spruit 2002]. A more realistic estimate of the rotational profile of the WD should take composition gradients into account.} We find
$t_{TS}\equiv \int_0^R(r/\nu_{TS})dr\approx 
10^{3}\,{\rm yr}\,(P/45{\rm min})^{3/2}$. 
Thus we expect the coupling time to lie in the range 
$t_{\rm coup} \lesssim 10^{3}\,{\rm yr}$ for the 
short orbital periods of interest.

\begin{figure}
\begin{centering}
\includegraphics[scale=.6]{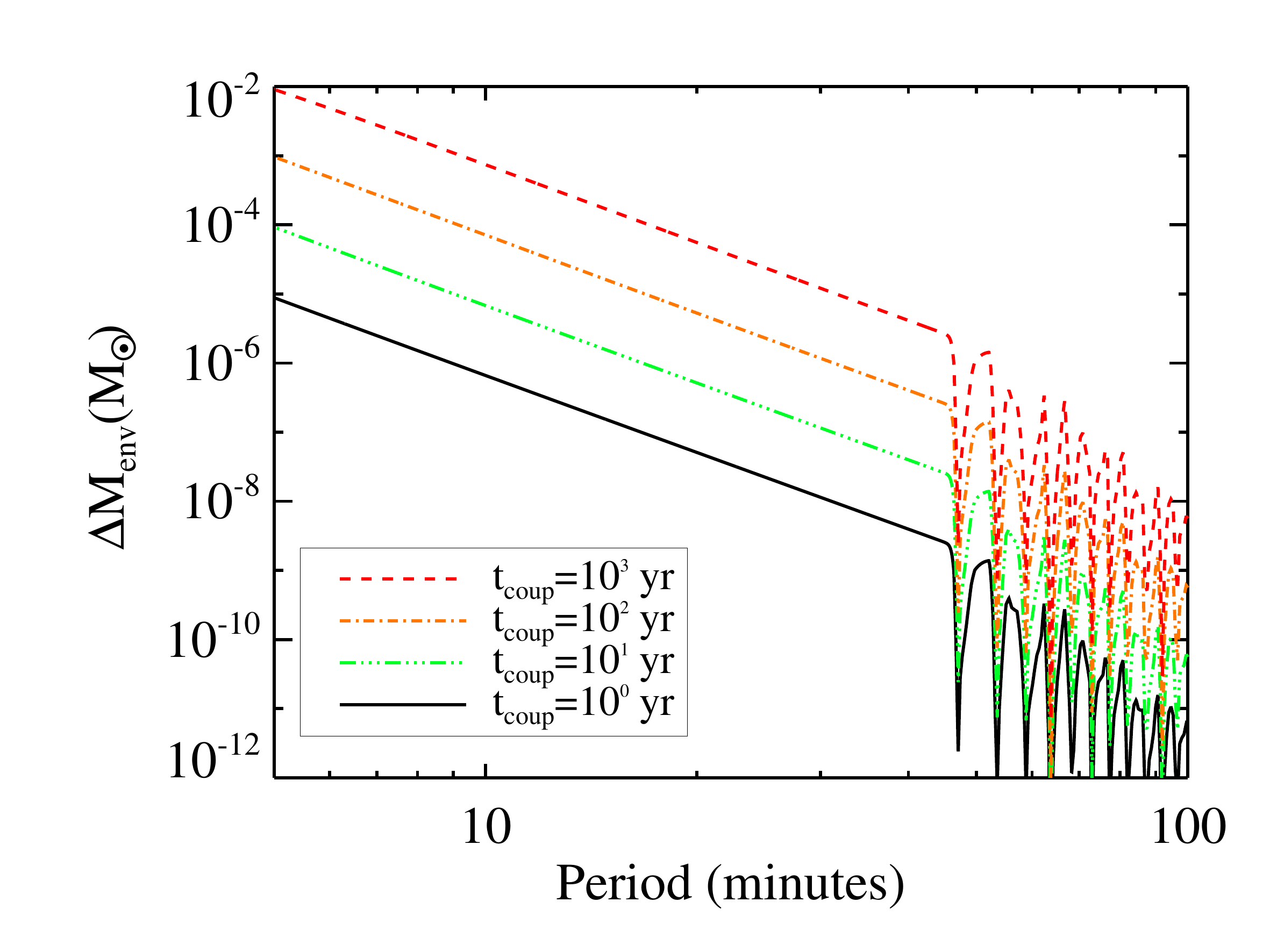}
\caption{\label{WDcrit} The mass $\Delta M_{\rm env}$ of the synchronized envelope
 as a function of orbital period for a $0.6
  M_\odot$ CO WD model with a $0.3 M_\odot$ companion. The solid (black)
  line has $t_{\rm coup} = 1\,{\rm yr}$, the dot-dot-dashed (green)
  line $t_{\rm coup} = 10\,{\rm yr}$, the dot-dashed (orange) line
  $t_{\rm coup} = 10^2\,{\rm yr}$, and the dashed (red) line
  $t_{\rm coup} = 10^3 {\rm yr}$.}
\end{centering}
\end{figure}

Figure \ref{WDcrit} plots the value of $\Delta M_{\rm env}$ as a
function of orbital period for our $0.6 M_\odot$ WD model with a $0.3
M_\odot$ companion, using values of $t_{\rm coup}$ ranging from $1
{\rm yr}$ to $10^3 {\rm yr}$. We begin our calculation at $P_{\rm orb}
> 1 {\rm hr}$ and use $I_{\rm env,0}=0$ and $\Omega_{{\rm core},0}=0$,
as appropriate at long orbital periods where tidal effects are
negligible. We see that for the range of $t_{\rm coup}$ considered, 
$\Delta M_{\rm env}$ remains small 
($\lesssim 10^{-2} M_{\odot}$) at all orbital periods of interest.
Thus, the synchronized envelope most likely does not extend down to the C/He
transition layer where the gravity waves are excited, justifying our 
assumption that $\dot{J}_z$ is a function of $\Omega_{\rm core}$.
However, the envelope does extend to very large optical depths,
suggesting that binary WDs may be observed to be synchronized at large
orbital periods even if their cores are not synchronized. Note that 
since $I_{\rm env} \ll I$, the core of the star contains most of the angular
momentum, and its spin evolves in the same manner as discussed in Paper
II.

\section{Tidal Heating and Unstable Nuclear Burning}

In the two-zone model discussed in \S 3, the total tidal heating rate 
$\dot E_{\rm heat}$ may be
calculated from equation (\ref{eheat}) with $\Omega_{s}=\Omega_{\rm
core}$, and the tidal heat is deposited entirely at the base of
the synchronized envelope where $\Delta M = \Delta M_{\rm env}$.
In a real WD, the heat deposition will occur over a range
of depths that depends on the details of wave breaking and viscous
angular momentum transport. For simplicity, here we choose to deposit the 
tidal heat uniformly per unit mass in the synchronized envelope.
The heating rate per unit mass, $\dot{\varepsilon}_{\rm heat}$, is then
\begin{align}
\label{epsheat1}
&\dot{\varepsilon}_{\rm heat} = 0 \quad {\rm for} \quad \Delta M > \Delta M_{\rm env} \\ \nonumber
&\dot{\varepsilon}_{\rm heat} = \frac{\dot{E}_{\rm heat}}{\Delta M_{\rm env}} \quad {\rm for} \quad \Delta M < \Delta M_{\rm env}.
\end{align}
Although the radial dependence of this heating function
is unlikely to be realistic, we find that the results below 
are not strongly dependent on the form of $\dot\varepsilon_{\rm heat}$.

To understand the effect of tidal heating on the WD properties, we
evolve WD models using the extra heating term calculated via equation
(\ref{epsheat1}). We use the one-dimensional stellar evolution code
MESA (Paxton et al.~2010) to evolve our WD models, starting from an initial
orbital period of one hour. We present results for a $0.6 M_\odot$ CO
WD model with a $\sim10^{-4} M_\odot$ hydrogen shell and a $0.3 M_\odot$
companion.

\begin{figure}
\begin{centering}
\includegraphics[scale=.6]{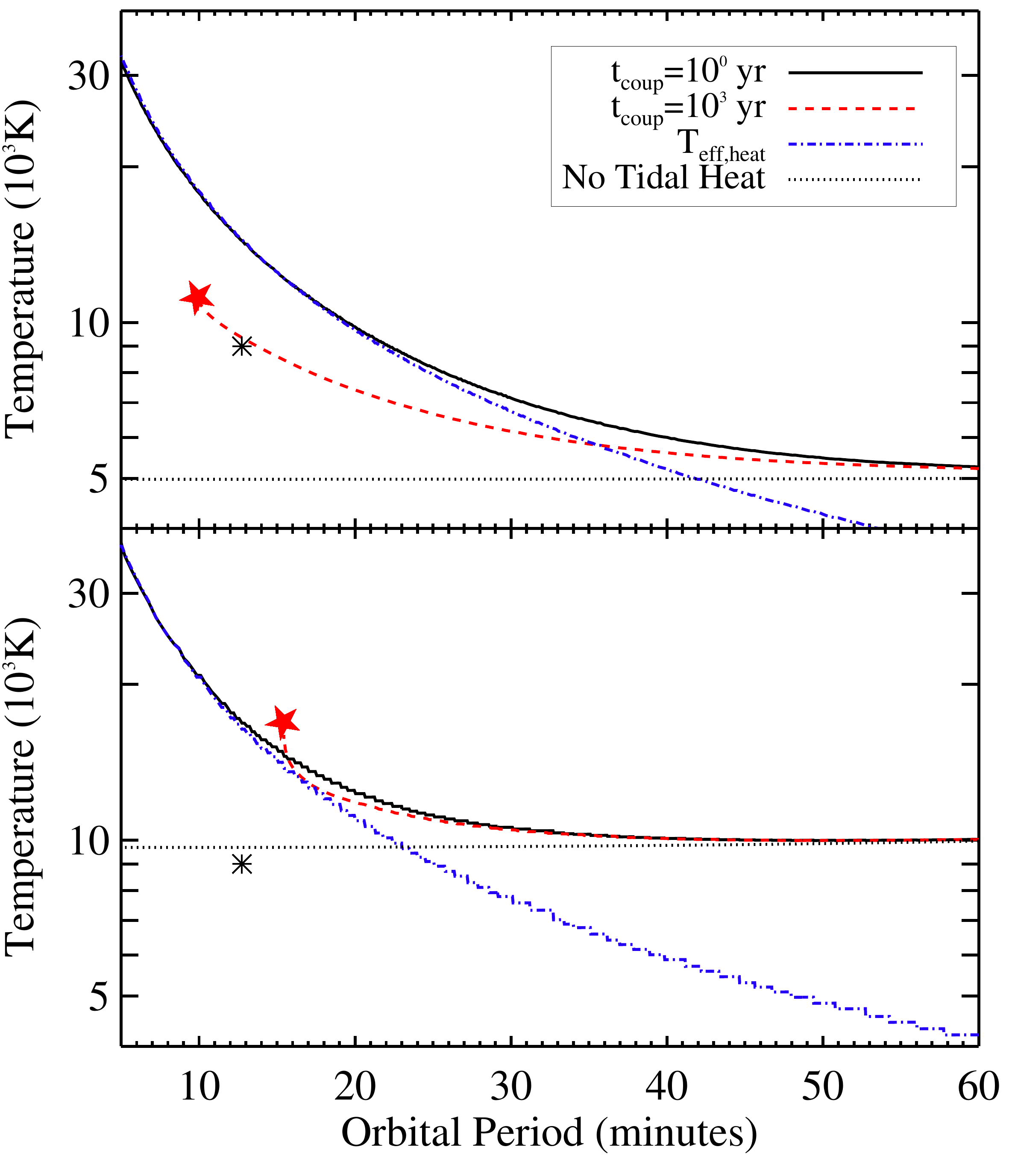}
\caption{\label{6temp} The surface temperature of the $0.6 M_\odot$ CO WD
  model with a $0.3 M_\odot$ companion as a function of orbital
  period, for initial temperatures of $5000$~K (top) and $10^4$~K
  (bottom). The solid black lines are calculated with $t_{\rm coup}=1\,{\rm yr}$ 
while the dashed (red) lines are calculated with $t_{\rm coup}=10^3\,{\rm yr}$. 
The dotted lines are calculated for a WD
  with no tidal heating and the same initial temperature. The (blue)
  dot-dashed lines correspond to equation
  (\ref{Ttide}). The (red) stars mark the
  points at which tidal novae occur.
The asterisks mark the position of the secondary of
the 12.75 minute binary WD system SDSS J065133+284423 (Brown et al.~2011).}
\end{centering}
\end{figure}

Figure \ref{6temp} displays the surface temperature as a function of
orbital period for our tidally heated WD. For comparison, we also show
the temperature of a non-tidally heated WD and the
``tidal heating temperature'', defined as
\be
\label{Ttide}
T_{\rm eff,heat}= \bigg(\frac{\dot{E}_{\rm heat}}{4\pi R^2 \sigma}\bigg)^{1/4}. 
\ee
At long orbital periods ($P \gtrsim 45$ minutes), the tidal heating
has little effect on the surface temperature of the WD. At
shorter periods ($P \lesssim 30$ minutes), the 
temperature becomes substantially larger due to tidal heating. 
Several of the curves end abruptly due to the ignition of a thermonuclear
runaway event, at which point we terminate our evolution calculations.

For small values of $t_{\rm coup}$, the tidal heat is deposited at
shallow depths and quickly diffuses to the surface such that the
luminosity of the WD is $L \simeq L_0 + \dot{E}_{\rm heat}$, where
$L_0$ is the luminosity of a non-tidally heated WD. 
However, for larger values of $t_{\rm coup}$, most of the tidal heat 
is deposited deeper in the WD where it cannot quickly diffuse outward. 
This leads to lower surface temperatures, although the
internal temperature may increase substantially.

\begin{figure}
\begin{centering}
\includegraphics[scale=.6]{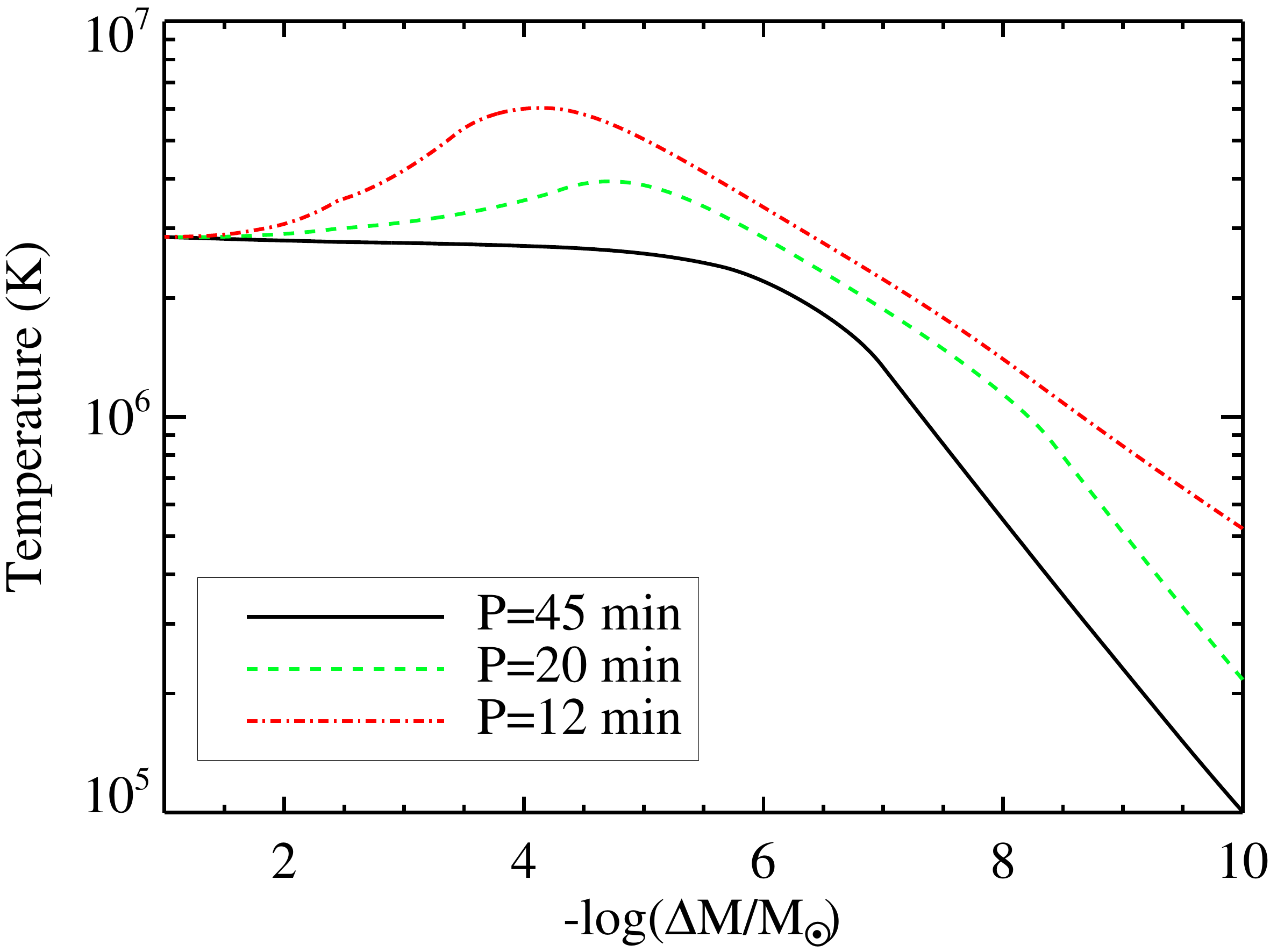}
\caption{\label{63temp} Temperature profile of the WD (as a function of exterior mass
  $\Delta M$) at orbital periods of 45 minutes (black), 20 minutes
  (green), and 12 minutes (red). These temperatures are calculated for
  the $0.6 M_\odot$ WD model with an initial surface temperature of $T_{\rm
    eff}=5000$~K, a $0.3 M_\odot$ companion and $t_{\rm coup}=10^3\,{\rm yr}$.}
\end{centering}
\end{figure}

Figure \ref{63temp} shows the interior temperature profile of 
our WD at three different orbital periods, using $t_{\rm coup}=10^3\,{\rm yr}$. 
At long orbital periods, the temperature profile is similar
to that of a non-tidally heated WD. As the orbital period decreases,
the interior heats up, with the local temperature maximum at 
$\Delta M\sim \Delta M_{\rm env}$. If the base of the hydrogen
layer reaches a temperature of $\sim 10^7$~K, hydrogen burning will be
ignited.

In the depicted model, the layer just above the He/H transition (at $\Delta M
\approx 10^{-4} M_\odot$) is composed of largely degenerate
hydrogen gas. The ignition of fusion in this layer can thus spark a
thermonuclear runaway.
In general, our calculations show that these {\it tidal novae} occur
only in initially cool WDs ($T_{\rm eff} \lesssim 1.2 \times 10^4$~K
in the absence of tidal heating). They do not occur in hotter WDs
because the hydrogen is not degenerate and can burn stably.
Also, tidal novae require that the waves
deposit some of the heat near the base of the hydrogen layer,
i.e., $10^{-5}M_\odot \lesssim \Delta M_{\rm env} \lesssim 10^{-3} M_\odot$. 
In our two zone model, such heating occurs for coupling times $10 {\rm yr} \lesssim t_{\rm coup} \lesssim
10^4 {\rm yr}$. Overall, we find that the tidal novae
occur at orbital periods $5~{\rm min}
\lesssim P_{\rm orb} \lesssim 20~{\rm min}$, depending on the location
of heat deposition, initial temperature of the WD, and companion mass.

\section{Discussion}

We have shown that under rather general conditions (see the last paragraph of Section 4), tidal dissipation in compact WD binaries can lead to nova outbursts prior to binary merger or mass transfer.
While we do not attempt to predict the detailed observational signal of a
tidal nova (TN), we speculate that it may be very similar to a
classic nova. However, in contrast to classical novae in CVs,
a TN would occur in a compact system with
no evidence for mass transfer. Our results indicate that a 
TN would precede the beginning of mass transfer or merger by
about $t_{\rm GW}/4 \sim 10^5-10^6$~yrs [see Eq.~(\ref{tgw})], provided the conditions outlined in the previous paragraph are satisfied.

In most classical novae, the initial outburst is followed by a period
of stable hydrogen burning at near the Eddington luminosity, in which
the hydrogen shell of the WD inflates to a radius of order
$R_\odot$. However, the ultracompact nature of the WD system involved
in a TN (where $a\sim R_\odot/4$) may preclude
such a phase because the stably burning hydrogen shell would inflate
beyond the WD's Roche lobe. This shell may then accrete on to the
companion star or be ejected from the system. Therefore, we expect
most of the hydrogen to be burned or ejected during in a TN.
In the absence of mass transfer to supply a fresh 
hydrogen, recurrent novae would be unlikely. Thus, the occurrence
rate of these TN may be comparable to that of WD mergers involving a CO WD.

Our theory can be constrained by comparing the prediction of our
tidal heating calculations to observed compact WD binaries. The 12.75~minute system 
SDSS JJ065133+284423 provides the best opportunity (Brown et al.~2011). This system is
composed of a primary with $T_{\rm eff}=16400$~K and mass $0.25 M_\odot$,
and a secondary with $T_{\rm eff}\approx 9000$~K and mass $0.55 M_\odot$.
Comparison with Figure \ref{6temp} indicates that the
luminosity of the secondary is likely dominated by tidal heating. Our
result for a CO WD with an initial temperature of $5000$~K and a value
of $t_{\rm coup}=10^3~{\rm yr}$ is most consistent with the 
observed temperature of the secondary. These results indicate that a
TN may occur in this system in the future.

In principle, tidal heating may change the structure of the WD enough
to alter the dynamics of gravity wave propagation. However, we find
that this is not the case (i.e., no interior convection zone forms),
with the exception of a thermonuclear runaway event. Our simple 
two-zone model for the WD obviously needs improvement, and 
we have neglected the effects of mixing induced by the breaking gravity waves
and viscous angular momentum transport. If the mixing is strong enough
to smooth out the WD composition gradients, the dynamics of gravity
wave excitation and tidal heat deposition may be 
altered. Furthermore, if the surface hydrogen mixes into the WD
interior where it burns, the surface hydrogen layer will be gradually depleted and a
TN will not occur. Observations of the ejecta of classical novae indicate
substantial enrichment with core elements, although the mixing
mechanism is not well understood (Truran 2002). 
These and other aspects of TN in compact WD binaries warrant further study.

Future observations may be able to test whether TN occur and in turn
provide information about the tidal processes at work in WD binaries. The observation
of a nova-like event in a system with no evidence for mass transfer would be strong
evidence for the existence of TN and for the tidal heating mechanism studied in this
paper. Measurements of hydrogen surface abundances in compact WD systems could
also constrain our theory. The observation of a WD with a thick hydrogen envelope
in a very tight ($P \lesssim 5$ minutes) detached binary would indicate TN do not usually occur. If
WDs in tight binaries are observed to have little to no hydrogen on their surface,
this may indicate that TN have stripped the surface hydrogen, or that the hydrogen
has been destroyed due to efficient mixing processes. Observations of compact binary
WDs detected in future surveys may provide opportunities to test these theories.


We thank Bill Paxton, Lars Bildsten, and Eliot Quataert for
useful discussions. JF acknowledges the hospitality (Fall 2011)
of the Kavli Institute for Theoretical Physics at UCSB (funded by
the NSF through Grant 11-Astro11F-0016) where part of the work was carried out.
This work has been supported in part by NSF grant AST-1008245, NASA grants
NNX12AF85G and NNX10AP19G.

\def\apj{{Astrophys. J.}}
\def\apjs{{Astrophys. J. Supp.}}
\def\mnras{{Mon. Not. R. Astr. Soc.}}
\def\prl{{Phys. Rev. Lett.}}
\def\prd{{Phys. Rev. D}}
\def\apjl{{Astrophys. J. Let.}}
\def\pasp{{Publ. Astr. Soc. Pacific}}
\def\aa{{Astr. Astr.}}
\def\aapr{{Astr. Astr. Rev.}}


\end{document}